\documentclass{aa}

\usepackage{txfonts}
\usepackage[ansinew]{inputenc}
\usepackage{textcomp}
\usepackage{graphicx}
\usepackage{latexsym}

\begin{document}

\title{K-band polarimetry of an Sgr~A* flare with a clear sub-flare structure\thanks{Based on observations at the Very Large Telescope (VLT) of the European Southern Observatory (ESO) on Paranal in Chile; Program: 077.B-0552(A), PI: A. Eckart.}}

\author{L. Meyer\inst{1}\thanks{Member of the International Max Planck Research School (IMPRS) for Radio and Infrared Astronomy at the Universities of Bonn and Cologne.} \and R. Sch\"odel\inst{1} \and A. Eckart\inst{1} \and V.~Karas\inst{2} \and M. Dov\v{c}iak\inst{2} \and W. J. Duschl\inst{3,4}  }

\institute{I.Physikalisches Institut, Universit\"at zu K\"oln, Z\"ulpicher Str. 77, 50937 K\"oln, Germany \and Astronomical Institute, Academy of Sciences, Bo\v{c}n\'{i} II, CZ-14131 Prague, Czech Republic \and Institut f\"ur Theoretische Physik und Astrophysik, Universit\"at zu Kiel, 24098 Kiel, Germany \and Steward Observatory, The University of Arizona, 933 N. Cherry Ave. Tucson, AZ 85721, USA }

\date{Received  / Accepted }

%%%%%
%
\abstract {The supermassive black hole at the Galactic center, Sgr~A*,
shows frequent radiation outbursts, often called 'flares'. In the
near-infrared some of these flares were reported as showing intrinsic
quasi-periodicities. The flux peaks associated with the quasi-periodic
behavior were found to be highly polarized.}
{The aim of this work is to present new evidence to
support previous findings of the properties of the polarized
radiation from Sgr~A* and to again provide strong support for the quasi-periodicity of $\sim$$18\pm 3$\,min reported
earlier.}
{Observations were carried out at the European Southern
Observatory's Very Large Telescope on Paranal, Chile. We used the
NAOS/CONICA adaptive optics/near-infrared camera instrument.  By
fitting the polarimetric lightcurves with a hot-spot model, we addressed
the question of whether the data are consistent with this model. To fit
the observed data we used a general relativistic ray-tracing code in
combination with a simple hot-spot/ring model.}
{We report on new polarization measurements of a K-band flare from the
supermassive black hole at the Galactic center. The data provide very strong
support for a quasi-periodicity of $15.5\pm2$\,min. The mean
polarization of the flare is consistent with the direction of the
electric field vector that was reported in previous observations. The
data can be modeled successfully with a combined blob/ring model. The inclination $i$ of the blob orbit must be $i > 20\degr$ on a 3$\sigma$ level, and the dimensionless spin parameter of the black
hole is derived to be $a_\star > 0.5$. }
{}


%\titlerunning{Constraints on the orbiting spot model for NIR Sgr~A* flares}

\maketitle

\section{Introduction}

The radio, infrared, and X-ray source Sagittarius~A* (Sgr~A*) at the
center of the Milky Way is generally accepted to be related to
emission by plasma in the immediate environment of a
$3.7\times10^{6}$\,M$_{\odot}$ black hole
(e.g. Schoedel et al.~\cite{rainer1}; Ghez et al.~\cite{ghez}; Eisenhauer et al.~\cite{eisenhauer}).
With Eddington luminosities of $L_{\rm Edd} = 10^{-9}-10^{-10}$, it is the most extreme sub-Eddington source accessible to
observations. A characteristic feature of Sgr~A* is the so-called
'flares', short bursts of increased radiation that last for about
60-100\,min. These flares were first discovered at X-ray wavelengths,
where the flux of Sgr~A* may rise by factors up to $\sim$100 during
such an event
(e.g. Baganoff et al.~\cite{baganoff}; Porquet et al.~\cite{porquet}; Eckart et al.~\cite{ecki3}).  At
near-infrared (NIR) wavelengths, the flares show very similar
timescales, but the flux varies only by factors of $\leq10$
(Genzel et al.~\cite{genzel}). Although the exact cause of the
flares is still unclear, it is generally accepted that they are caused
by synchrotron and synchrotron self-Compton emission processes within
$\la 10$\,Schwarzschild radii ($\mathrm R_{S}$) of the black hole
(Eckart et al.~\cite{ecki1}; Gillessen et al.~\cite{gillessen}). The non-thermal nature of the
flares has recently been shown directly by detecting polarized NIR
radiation from Sgr~A* (Eckart et al.~\cite{ecki2}).

The most intriguing feature related to these flares is quasi-periodic
oscillations (QPOs) with a period of 17-22\,min, which have been
detected in several of these events (Genzel et al.~\cite{genzel}; Belanger et
al.~\cite{belanger}; Eckart et al.~\cite{ecki2}; Meyer et al.~\cite{ich}). These periodicities may be related to the
high-frequency quasi-periodic oscillations (HFQPO) observed in some
black hole binaries. These HFQPOs scale inversely with the root of the
mass and are thought to be related to plasma in a relativistic flow
within a few Schwarzschild radii of the black hole. Although their
mechanism has still not been clearly understood, they appear to be
promising tools in probing the space time around black holes (for a
review of black hole binaries and HFQPOs, see Nowak \& Lehr~\cite{nowak}
; McClintock \& Remillard~\cite{mcclintock}). The QPOs
observed in Sgr~A* are similar to HFQPOs in the sense that they appear
to fit into the mass-scaling relationship. However, the exact relation
between the XRB HFQPOs and Sgr~A* QPOs is not clear yet. In particular,
XRB HFQPOs are observed at X-ray wavelengths, arise in thin accretion
disks, and show amplitude modulations of just a few percent. No
stationary thin disk is thought to be present in Sgr~A*, the QPOs are
observed at NIR wavelengths, and they show a modulation of $>10\%$ (up
to 50\% as reported in this paper). 

The QPOs may provide the possibility of measuring the spin of the black
hole (see, e.g., Genzel et al.~\cite{genzel}; Broderick \& Loeb~\cite{broderick1}; Meyer
et al.~\cite{ich}). However, these QPOs have only been observed
unambiguously in a few cases, which has raised some doubts about their
nature, particularly whether they may just be related to a
red-noise-like process in the source. This aspect is discussed
extensively in Meyer et al.~(\cite{ich}) for the flare that shows the clearest evidence of quasi-periodicities. They
conclude that the $\sim$17\,min periodicity reported by
Genzel et al.~(\cite{genzel}) is most probably not due to red noise. The
discovery by Eckart et al.~(\cite{ecki2}) that periodicities can
show up in NIR polarized light, while at the same time being hardly
visible in the total flux, raises the possibility that such
periodicities may be present in most flares. However, due to the
difficulties in NIR polarization measurements (e.g. need for
excellent seeing, need for rapid change of polarization angles due to
intrinsic variability of Sgr~A*), this observing window has only been opened
recently. So far, only two flares have been reported in NIR
polarized light (Eckart et al.~\cite{ecki2}).

Here we report on the recent NIR  polarimetric observation  of a new
flare from Sgr~A*. The flare was exceptionally bright and shows
intrinsic variability on a level of $\geq50\%$.

\section{Observation and data reduction}  

Using the NIR camera CONICA and the adaptive optics (AO) module NAOS
on ESO's Very Large Telescope UT4 on Paranal in Chile, we observed
Sgr~A* in the K$_S$-band during the night between 31 May and 1 June
2006. To achieve good time resolution, a Wollaston prism was combined
with a half-wave retarder plate. This allows the simultaneous
measurement of two orthogonal directions of the electric field vector
and a rapid change between different angles.The detector integration
time was 30\,s. Including the overheads due to the need to turn the
half-wave plate and telescope offsets, individual frames could be
taken every 70-80\,s.

During the observation, the optical seeing ranged between 0.6\arcsec
and 1\arcsec and the AO correction was stable. Sky measurements were
taken by observing a dark cloud a few arc-minutes to the north west of
Sgr~A*. To minimize the effects of dead pixels, the observations were
dithered. The data were reduced in a standard way, i.e. sky
subtracted, flat-fielded, and corrected for bad pixels. For every
individual image, the point spread function (PSF) was extracted
with the code {\it StarFinder} by Diolaiti et al.~(\cite{diolaiti}). Each
exposure was deconvolved with a Lucy-Richard deconvolution and restored with a Gaussian beam. The flux of Sgr~A* and other compact sources in the field was obtained via aperture photometry on the
diffraction-limited images. The background flux density was determined
as the mean flux measured with apertures of the same size at five
different positions in a field that shows no individual stars.
Photometric calibration was done relative to stars in the field with a
known flux. For the extinction correction we assumed
$A_K=2.8$\,mag. Estimates of uncertainties were obtained from the
standard deviation of fluxes of nearby constant sources. The
calibration was performed using the overall interstellar polarization
of all sources in the field, which is 4\% at $25\deg$
(Eckart et al.~\cite{ecki95}; Ott et al.~\cite{ott}).

\begin{figure}
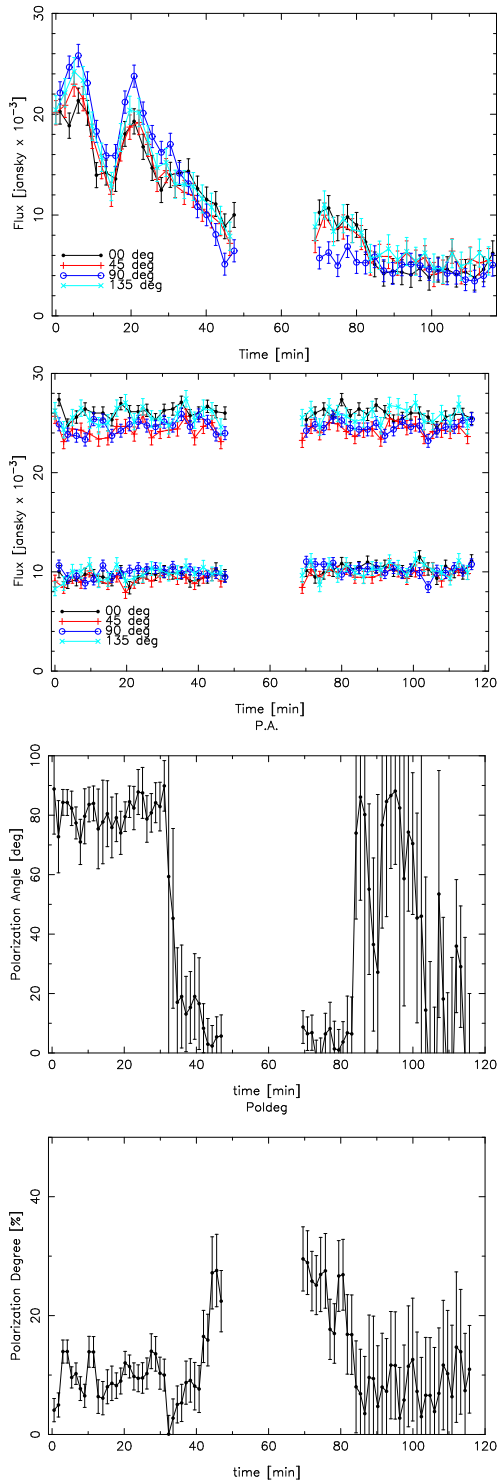

\centering 
\includegraphics[angle=270,width=6.5cm]{pol2006_flux_calib_sgr.eps}
\includegraphics[angle=270,width=6.5cm]{pol2006_flux_calib_comp.eps}
\includegraphics[angle=270,width=6.5cm]{PA_31May2006.eps}
\includegraphics[angle=270,width=6.5cm]{Poldeg_31May2006.eps}

\caption{Dereddened flux of Sgr~A* (top) and of the two comparison
stars W6 and S7 (second from top) at the different polarization angles. The
comparison stars are located within $<1\arcsec$ of Sgr~A*. Due to the
presence of two stellar sources at the position of Sgr~A* (see Fig.~1 in
Eisenhauer et al.~\cite{eisenhauer}), its lightcurve stays at a constant
level of about 5\,mJy after the flare. The light curves of the
channels have been calibrated to an overall polarization of $4\%$ at
an angle $25^\circ$ east of north
(Eckart et al.~\cite{ecki95}; Ott et al.~\cite{ott}). Also, the curves are shifted to the
\emph{total} flux of the sources, i.e. the flux per channel is only
half of the flux shown in the plots. The inferred polarization angle and the degree of linear polarization of Sgr~A* can be seen in the bottom panels. \label{Fig:lightcurve}}
\end{figure}

Fig.~\ref{Fig:lightcurve} (top) shows the dereddened light curve of Sgr~A*
 for all four measured polarization
angles. Please note that the flux was calibrated relative to stars of
known flux in the field-of-view for each angle separately. Therefore,
the mean of the shown lightcurves is as high as the total flux of a
source, i.e., actually the figure shows twice the flux for each angle
(the same convention was used in Eckart et al.~\cite{ecki2}).

\section{Observed variability of Sgr~A*}

The light curve shown in Fig.~\ref{Fig:lightcurve} shows that two
peaks, i.e. sub-flares, can be clearly distinguished at all four
polarization angles. In contrast to the data of 2005 (Eckart et al.~\cite{ecki2}), these two sub-flares show up clearly in each polarization
channel.

\begin{figure}
\centering
\includegraphics[angle=270,width=6.5cm]{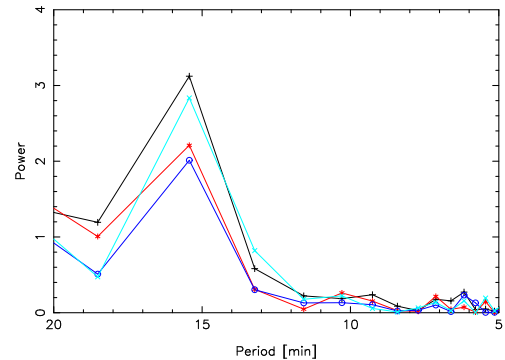}
\caption{Periodogram of the light curves of Sgr~A* for all four
  observed polarization angles. The periodogram has been oversampled
  by a factor of 2.
\label{Fig:period}}
\end{figure}

Fig.~\ref{Fig:period} shows a periodogram of the emission from
Sgr~A* for the first 50\,min of the flare. For clarity, the
periodogram has been oversampled by a factor of two. A clear peak is
present at a period of $15.5\pm2$\,min. There is no indication of red
noise. This was also checked by plotting a log-log version
of the periodogram, where red-noise would show up as a straight line
with slope one. It is certainly statistically questionable to infer
quasi-periodicity from just two peaks. However,
considering that very similar periodicities have already been reported for
several NIR flares clearly changes the picture.  The
probability of this happening in the case of pure red noise is extremely
low (see discussion in Meyer et al.~\cite{ich}).

The Stokes parameters $I$, $Q$, and $U$ were obtained by measuring the
flux at position angles (PA) of the electric field vector of $0\degr$,
$45\degr$, $90\degr$, and $135\degr$ (E of N).  The total flux, the
polarization angle, and polarization degree can be inferred from the single polarization
channels and are also shown in Fig.~\ref{Fig:lightcurve} (lower two panels). The PA of
$80\degr \pm 10\degr$ during the sub-flares agrees very well with the
value of $60\degr \pm 20\degr$ reported by Eckart et al.~(\cite{ecki2}). This is a strong indication of a stable arrangement of the
accretion flow with respect to the spin axis of the MBH, which may
point to a permanent accretion ring with a radial extent of
$\sim$\,2\,$R_S$ (see also Meyer et al.~\cite{ich};
Moscibrodska et al.~\cite{moscibrodzka}). Therefore the flow is
probably not fully advective very close to the BH horizon. Since the
accretion over a disk is more effective than that of an advective
flow, solutions within radiatively inefficient accretion flows (RIAFs)
might be preferred where $\dot{M}$ is small and the outflow is large.

A striking feature is the large amplitude of the two sub-flares. It
reaches about 50\% of the overall flare intensity. The observed
frequencies of QPOs in X-ray binaries scale with the BH mass, which
can be extrapolated to the mass of Sgr~A*. Therefore the large
amplitude is remarkable since the amplitude of high-frequency QPOs
observed for X-ray binaries shows variability amplitudes not larger
than a few percent (see reviews by Nowak \& Lehr~\cite{nowak}; McClintock \& Remillard~\cite{mcclintock}).

The lightcurves show an interesting feature between $\sim$\,40\,--\,85\,min. The polarization degree rises sharply up to $\sim$\,30\% and the PA swings to 0\degr  (see Fig.~\ref{Fig:lightcurve}). It is not clear whether this is an effect intrinsic to Sgr~A*. The feature occurs after the bright flare phase, when Sgr~A* is quiescent, i.e. in a low flux state. Therefore the polarization measurements are less certain. On the other hand, this feature shows some robustness, as it is visible before and after the sky observations. Also, the FWHM of the PSFs that have been extracted from each image shows no conspicuous behavior during that time, which means that the AO correction is most probably not responsible for this property of the lightcurves. It could be possible that this effect is a flare of Sgr~A* that is mainly visible in polarization. A similar observation was recently made by Eckart et al.~(\cite{ecki2}), who found that some sub-flares may only show up in the polarized flux.

\section{A plasma blob close to the BH horizon} 

In this section we show that the observed polarimetric light curves
are in excellent agreement with a combined hot spot/ring model. In
this model the sub-flares are due to a blob on a relativistic orbit
around the MBH, while an underlying ring accounts for the broad overall
flare. Relativistic effects like beaming, lensing, and change of
polarization angle imprint on the emitted intrinsic radiation
(e.g. Dovciak et al.~\cite{dovciak}; Connors \& Stark~\cite{connors}; Hollywood \& Melia~\cite{hollywood2}; Broderick \& Loeb~\cite{broderick2}). In our model we assume that the variability
in the polarization angle and the polarization degree are only due to the
relativistic effects. As the emitted radiation of Sgr~A* is
synchrotron radiation (emitted in the disk corona), we assumed two different magnetic field
configurations to fit the light curves with our model. The first is
analogous to a sunspot and results in a constant $E$-vector
perpendicular to the disk. The second configuration is a global
azimuthal magnetic field that leads to a rotation of the $E$-vector
along the orbit. More details on the model and the fitting procedure
can be found in Meyer et al.~(\cite{ich}).

\begin{figure}[t]
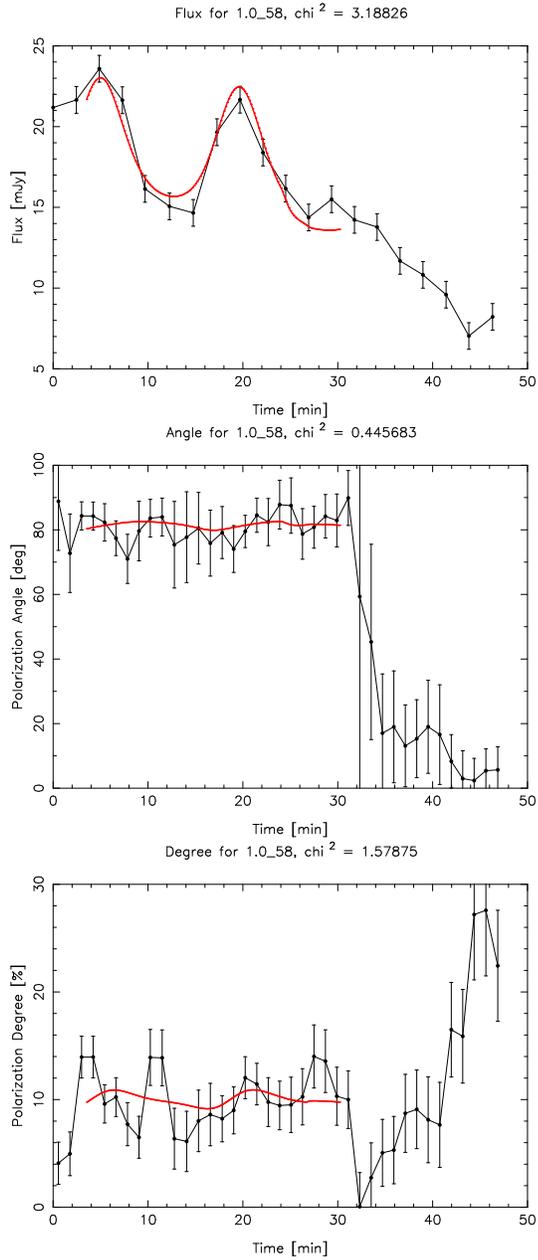

\centering
\resizebox{7cm}{!}{\includegraphics[angle=270]{1.0_58_2006.eps}}

\resizebox{7cm}{!}{\includegraphics[angle=270]{angle_1.0_58_2006.eps}} 

\resizebox{7cm}{!}{\includegraphics[angle=270]{poldeg_1.0_58_2006.eps}}
\caption{The best fit solution (in red) for the constant $E$-vector case. Shown is the flux (top), the polarization angle (middle), and the degree of linear polarization (bottom). The parameters of the model are $i = 58\degr$, $a_\star \approx 1$. The spot is orbiting at a \mbox{radius $r=4\,GM/c^2$.} The disk/ring and the spot both have an intrinsic polarization degree of $\sim$\,17\%.}
\label{fits}
\end{figure}

The fit with the least reduced-$\chi^2$ value is shown in
Fig.~\ref{fits}. A dimensionless spin parameter of $a_\star \approx
1$ and an inclination of $i \approx 58\degr$ give the best
solution. Only the time interval $\leq 30$\,min was fitted
because the spot emission disappears afterwards. The magnetic
field corresponds to the constant $E$-vector scenario, which gives
better fits than the azimuthal-field case. It is also interesting that
the best-fit values for the intrinsic polarization degree of the disk/ring and the spot are the same ($\sim$\,17\%). This is a profound difference to the data presented by Eckart et al.~(\cite{ecki2}) where the spot was intrinsically highly polarized ($\sim$\,50\%), while the disk was unpolarized ($\sim$\,5\%).   

\begin{figure}[t]
\centering
\resizebox{7cm}{!}{\includegraphics{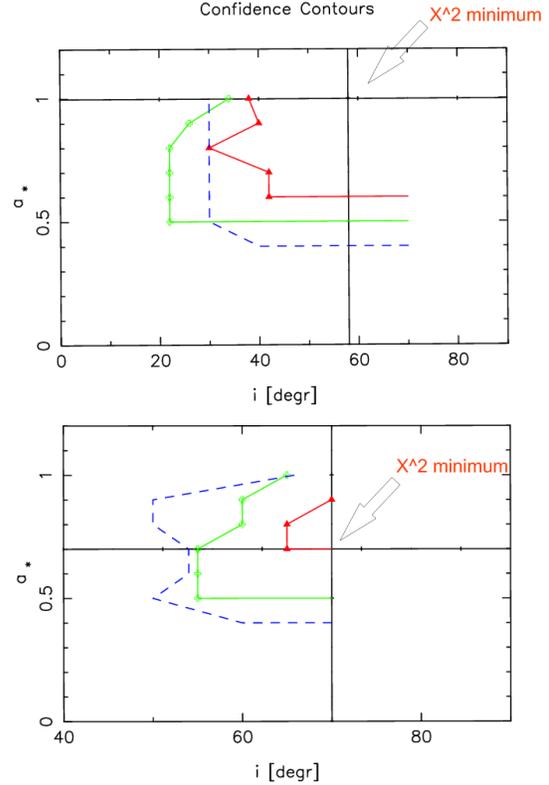}}

\caption{Confidence contours for the constant $E$-Vector case (top) and the azimuthal magnetic field (bottom). The red (green) lines are chosen such that the projection onto one of the parameter axes gives the 1$\sigma$ (3$\sigma$) limit for this parameter. The blue dashed lines indicate the 3$\sigma$ contour for the 2005 data, analyzed by Meyer et al.~(\cite{ich}). The $\chi^2$-minimum for the 2006 data is marked by the big cross. The $\chi^2$-minimum of the 2005 epoch lies at $a_\star = 1$, $i=70\degr$ (top) and $a_\star = 0.5$, $i=70\degr$ (bottom). Our analysis is limited to $i \lessapprox 70^\circ $, see Meyer et al.~(\cite{ich}).}
\label{conf}
\end{figure}

As noted by Meyer et al.~(\cite{ich}), a property of the
model is a relatively weak dependence on the spin parameter. Figure~\ref{conf} (top) shows the confidence contours for the constant $E$-Vector case within the $a_\star$-$i$-plane. The spin parameter can only be constrained to the region $a_\star \ga 0.5$ on a 3$\sigma$ level. Note that the observed timescale of roughly 15\,min means that the spot is inside the least stable orbit, i.e. freely falling, for $a_\star \la 0.6$. The inclination is $i > 20\degr$ on a 3$\sigma$ level. The confidence contours for the azimuthal magnetic field are shown in Fig.~\ref{conf} (bottom). The minimum $\chi^2$ value here is $1.2$ higher than in the upper case, indicating a worse fit. The inclination can be constrained to $i \geq 55\degr$ on a $3\sigma$ level.

Our results agree very well with Eckart et al.~(\cite{ecki2}) and
Meyer et al.~(\cite{ich}) who analyzed polarimetric data from
2005 and found least $\chi^2$-values for high spin parameters
($a_\star \approx 1$) and high inclinations ($i \approx
70\degr$), but with the same characteristic of a weak
dependence, see blue dashed lines in Fig.~\ref{conf}. 
To progress in the descritpion of the Sgr~A* system, bright, strongly polarized sub-flares and a faint, low-polarized disk emission have to be observed. Accurate data of this type not only could give tighter constraints on the inclination and the spin parameter but also test the Kerr metric qualitatively. Due to slightly different observed frequencies of the different
epochs, a hot spot has to orbit on a
slightly different orbit. But this should lead to the same spin
parameter, as the mass of the BH is not expected to change
significantly.

Although it
would actually be difficult to explain why a single blob survives that
long in the place where shearing is enormous, the
repeated observation of 16--20 min separation among individual
sub-flares justifies our assumption that assigns the same confined region to
different peaks. This empirical indication can be taken as a constraint
that needs to be imposed on theoretical models for the origin and confinement of blobs
and the onset of flares in Sgr A*.

{ While the observation of QPOs seems to favor the orbiting spot model instead of an adiabatically expanding plasma blob, the model of relativistic plasma clouds expanding in a cone-jet geometry, in addition to suitable instabilities within the jet, can account for the flares and sub-flares of Sgr~A* as well (Yusef-Zadeh et al.~\cite{yusef}; Eckart et al.~\cite{ecki1}). In fact, the inclusion of both models offers a possibility of linking the NIR/X-ray activity to the sub-mm/radio regime: a plasma blob that is orbiting very close to the BH horizon (as described above) and then adiabatically expanding along a short jet-like geometry or within its orbit may explain the observed variability of Sgr~A* across all wavelengths. }

\begin{acknowledgements}
We are grateful to C. Kiefer. This work was supported in part by the Deutsche Forschungsgemeinschaft (DFG) via grant SFB 494. We want to thank all members of the NAOS/CONICA and the ESO PARANAL team, especially N. Ageorges and M. Messineo for support in setting up the polarization measurements.
\end{acknowledgements}

%\bibliography{gc}

\end{document}